\documentstyle[aps]{revtex}
\begin{document}
\bibliographystyle{prsty}
\baselineskip=8.5mm
\parindent=7mm
\begin{center}
{\Huge {\bf \sc{ decay constants of the pion and B mesons with the Bethe-Salpeter equation}}} \\[2mm]
Zhi-Gang Wang$^{1,2}$ \footnote{E-mail,wangzgyiti@yahoo.com.cn}, Wei-Min Yang$^{2,3}$ and Shao-Long Wan$^{2,3} $    \\
$^{1}$ Department of Physics, North China Electric Power University, Baoding 071003, P. R. China \footnote{Mailing address}\\
$^{2}$ CCAST (World Laboratory), P.O.Box 8730, Beijing 100080,
P.R.China \\
$^{3}$ Department of Modern Physics, University of Science and Technology of China, Hefei 2300026, P. R. China \\
\end{center}

\begin{abstract}
In this article, we investigate the under-structures of the $\pi$ and B mesons in the framework of
the Bethe-Salpeter equation with the confining effective potential (infrared modified flat bottom potential).
In bare quark-gluon vertex approximation, we obtain the
algebraic expressions for the solutions of the coupled rain-bow Schwinger-Dyson equation and ladder
Bethe-Salpeter equation.    Firstly, we neglect the rain-bow Schwinger-Dyson equation,
 take the bare quark propagator and solve the Bethe-Salpeter equation numerically alone.
 Although the bare quark propagator can not embody dynamical chiral symmetry breaking and
has a mass pole in the time-like region, it can give reasonable results for the values of
decay constants $f_\pi$ and $f_B$ compared with the values of experimental data and other theoretical calculations,
such as lattice simulations  and QCD sum rules. Secondly, we explore those mesons within the framework of
the coupled rain-bow Schwinger-Dyson equation and ladder
Bethe-Salpeter equation.
The Schwinger-Dyson functions for the $u$ and $d$ quarks are greatly renormalized  at small momentum region
and the curves are steep at about $q^2=1GeV^2$ which indicates an explicitly dynamical symmetry breaking.
The Euclidean time fourier transformed quark propagator has no mass poles in the time-like region which
 naturally implements  confinement.
As for the $b$ quark, the current mass is very large, the renormalization is more tender, however,
 mass pole in the time-like region is also absent. The Bethe-Salpeter wavefunctions for both
the $\pi$ and B mesons have the same type (Gaussian type) momentum dependence as the corresponding wavefunctions
with the bare quark propagator,
 however, the quantitative values are changed and the  values for the decay constants $f_\pi$ and $f_B$ are changed correspondingly.
\end{abstract}


\section{Introduction}
 Quantum chromodynamics (QCD) is the appropriate theory for describing the strong interaction at high
energy region, however, the strong gauge coupling at low energy destroys the perturbative expansion  method.
The physicists propose many nonperturbative approaches
to deal with the long distance properties of QCD, such as Chiral perturbation theory \cite{Gasser85},
heavy quark effective theory \cite{Neubert94}, QCD sum rule \cite{Shifman79}, lattice QCD \cite{Gupta98},
perturbative QCD \cite{Brodsky80}, coupled Schwinger-Dyson equation (SDE) and
Bethe-Salpeter equation (BSE) method \cite{Roberts94}, etc. All of those approaches have both outstanding advantages
and  obvious shortcomings. For example, lattice simulations are rigorous in view of QCD, they suffer
from lattice artifacts and uncertainties, such as Gribov copies, boundary conditions and so on, furthermore
current technique can not give reliable result below 1 GeV, where the most interesting and novel behavior
is expected to lie. The coupled SDE and BSE have given a lot of successful
descriptions of the long distance properties of strong interactions and the QCD vacuum, for a recent review one
can see Ref.\cite{Roberts03}. The SDE can provide a natural way to embody the dynamical symmetry breaking and confinement
which are two crucial features of QCD, although they correspond to two very different energy scales \cite{Miransky93,Alkofer03}.
On the other hand, the BSE is a conventional approach in dealing with the two body
relativistic bound state problems \cite{BS51}. From the solutions of the BSE, we can obtain useful information
about the under-structure of the hadrons and thus obtain powerful tests for the quark theory of the mesons.
However, the main drawback can be traced back to the fact that when we solving
the SDE and BSE, model dependent kernels for the gluon two point Green's function have to be used,
furthermore, the coupled SDE and BSE are a divergent series of equations, we have to make
truncations in one or the other ways. Numerical calculations indicate that
the coupled rainbow SDE and ladder BSE with phenomenological potential models can give satisfactory results. The usually used
effective potentials are confining Dirac $\delta$ function potential,
Gaussian  distribution potential and flat bottom potential (FBP) \cite{Munczek91,SC,Wangkl93}. The FBP is a sum of Yukawa
potentials, which not only  satisfies gauge invariance, chiral invariance and fully
relativistic covariance, but also suppresses the singular point which the
 Yukawa potential has. It works well in understanding the dynamical chiral symmetry braking, confinement and the QCD vacuum as well as
the meson  structures, such as electromagnetic form factor, radius, decay constant \cite{WangWan,Wan96}.

The decay constant of the B meson $f_B$ plays an important role in modern physics with the assumption
of current-meson duality. The precise knowledge of the value of the $f_B$ will provide great  improvement in our
understanding of various processes convolving the B meson decays. At present, it is a great challenge to extract
the value  of the B meson decay constant $f_B$ from experimental data. So it is interesting to combine the those
successful  potentials within the framework of coupled SDE and BSE to calculate the decay constants of both
the $\pi$ and B mesons. In this article, we use an infrared modified flat-bottom potential (IMFBP) which takes
 the advantages of
both the Gaussian distribution potential and the FBP to calculate both the $\pi$ and
B mesons decay constants. Certainly, our potential model can be used to investigate  the properties of
other pseudoscalar mesons, such as $K,D,D_s \cdots$. For example, we can obtain
the decay constants
$f_{\pi}=127 MeV, \ \ f_{K}=156 MeV, \ \ f_{D}=238 MeV$ and $f_{B}=192 MeV$ with the same parameters,
 while a detailed studies of those mesons $K,D,D_s \cdots$ may be our next work, they are not our main concern in
this article.

The article is arranged as follows:  we introduce the infrared modified flat bottom potential in section II;
in section III and IV, we solve the Schwinger-Dyson equation and the Bethe-Salpeter equation
and obtain the decay constants for both the $\pi$ and B mesons; section V is reserved for conclusion and discussion.
\section{Infrared modified Flat Bottom Potential }
The infrared structure of the gluon propagator has important
implication for the  quark confinement. One might expect that the behavior
of the quark interaction in the region of small  space-like $p^2$
determines the long range properties of the $q\bar{q}$ potential and hence
implements  confinement, however, the present techniques in QCD manipulation can not
give satisfactory small $r$ behavior for the gluon propagator, on the other hand, the phenomenological confining
 potential models give a lot of successes in dealing with the low energy hadron physics,
such as dynamical chiral symmetry braking, pseudoscalar mesons electromagnetic form factors,
mass formulations, $\pi-\pi$ scattering parameters, etc \cite{Roberts03,Munczek91,Tandy97}.
In this article, we use
a gaussian distribution function to represent the infrared behavior of the gluon propagator,
\begin{eqnarray}
4\pi G(k^2)=3\pi^2 \frac{\varpi^2}{\Delta^2}e^{-\frac{k^2}{\Delta}},
\end{eqnarray}
which determines the quark-quark interaction through a strength parameter $\varpi$ and a ranger parameter $\Delta$ \footnote{Here we correct a writing error in the first version.}.
This form is inspired by the $\delta$ function potential (in other words the infrared dominated potential) used in Ref.\cite{Munczek91}, which it approaches in the limit
$\Delta\rightarrow 0$. For the intermediate momentum, we take the FBP as the best approximation and neglect
the large momentum contributions from the perturbative QCD calculations as the coupling constant at high energy
is very small.
The FBP is a sum of Yukawa potentials which is an analogous to the
exchange of a series of particles and ghosts with different
masses (Euclidean Form),
\begin{equation}
G(k^{2})=\sum_{j=0}^{n}
 \frac{a_{j}}{k^{2}+(N+j \rho)^{2}}  ,
\end{equation}
where $N$ stands for the minimum value of the masses, $\rho$ is their mass
difference, and $a_{j}$ is their relative coupling constant.

The definition of momentum regions between infrared and intermediate momentum is about $\Lambda_{QCD}=200MeV$,
which is naturally set up by the minimum value of the masses $N=1\Lambda_{QCD}$, where the gaussian function
$e^{-\frac{k^2}{\Delta}}$ decays to about $0.3$ of its original values. Certainly, there are some overlaps
between those regions, in this way, we can guarantee the continuity  of the momentum. The asymptotic freedom
tell us that at high energy the gauge coupling is very  small and can be neglected safely, on the other hand,
 our phenomenological potential at energy
about $N+j\rho, j>3$ is already extend to the perturbative region and catches some perturbative physical effects.
Thus, our phenomenological infrared modified FBP is supposed to embody  a great deal  of  physical information
about all the momentum regions.

 Due to the particular condition we take for the FBP,
there is no divergence in solving the SDE.
In its three dimensional form, the FBP takes the following form:
\begin{equation}
V(r)=-\sum_{j=0}^{n}a_{j}\frac{{\rm e}^{-(N+j \rho)r}}{r}  .
\end{equation}
In order to suppress the singular point at $r=0$, we take the
following conditions:
\begin{eqnarray}
V(0)=constant, \nonumber \\
\frac{dV(0)}{dr}=\frac{d^{2}V(0)}{dr^{2}}=\cdot \cdot
\cdot=\frac{d^{n}V(0)} {dr^{n}}=0    .
\end{eqnarray}
So we can determine $a_{j}$ by solve the following
equations, inferred from the flat bottom condition Eq.(4),
\begin{eqnarray}
\sum_{j=0}^{n}a_{j}=0,\nonumber \\
\sum_{j=0}^{n}a_{j}(N+j \rho)=V(0),\nonumber \\
\sum_{j=0}^{n}a_{j}(N+j \rho)^{2}=0,\nonumber \\
\cdots \nonumber \\
\sum_{j=0}^{n}a_{j}(N+j \rho)^{n}=0 .
\end{eqnarray}
As in  previous literature \cite{Wangkl93,WangWan,Wan96}, $n$ is set to be 9.
\section{Schwinger-Dyson equation}
The Schwinger-Dyson equation, in effect the functional
 Euler-Lagrange equation of the quantum field theory, provides a natural
  framework for investigating the nonperturbative properties  of the
  quark and gluon Green's functions. By studying the evolution
  behavior and analytic structure of the dressed quark propagator,
  one can obtain valuable information about the dynamical symmetry
  breaking phenomenon and confinement.
 The SDE for the quark takes the following form:
\begin{equation}
S^{-1}(p)=i\gamma \cdot p + m+\frac{16 \pi i}{3}\int
\frac {d^{4}k}{(2 \pi)^{4}} \Gamma_{\mu}
S(k)\gamma_{\nu}G_{\mu \nu}(k-p),
\end{equation}
where
\begin{eqnarray}
S^{-1}(p)&=& i A(p^2)\gamma \cdot p+B(p^2)\equiv A(p^2)
[i\gamma \cdot p+m(p^2)], \\
G_{\mu \nu }(k)&=&(\delta_{\mu \nu}-\frac{k_{\mu}k_{\nu}}{k^2})G(k^2),
\end{eqnarray}
and $m$ stands for an explicit quark mass-breaking term. In the rainbow approximation, we take $\Gamma_\mu=\gamma_\mu$.
With the explicit small mass term for the $u$ and $d$ quarks, we can preclude the zero
solution for the $B(p)$ and in fact there indeed exists a small bare current
quark mass. In this article, we take Landau gauge.
This dressing comprises the notation of constituent quark by
 providing a mass $m(p^2)=B(p^2)/A(p^2)$, which is corresponding to
 the dynamical symmetry breaking. Because the form of
 the gluon propagator $G(p)$ in the infrared region can not be exactly inferred from the $SU(3)$ color gauge theory,
 one often uses model dependent forms as input parameters in the previous studies
  of the rainbow SDE
  \cite{Roberts94,Roberts03,Wangkl93,WangWan,Wan96,Tandy97}, in this article we use the infrared modified FBP
to substitute for the gluon propagator.

In this article, we assume that a Wick rotation to Euclidean variables is
allowed, and perform a rotation analytically continuing $p$ and $k$
into the Euclidean region where them can be denoted by $\bar{p}$ and $\bar{k}$,
 respectively.   Alternatively, one can derive the SDE from the
 Euclidean path-integral formulation of the theory, thus avoiding
 possible difficulties in performing the Wick
 rotation $\cite{Stainsby}$ . As far as only numerical results are concerned,
  the two procedures are equal. In fact, the analytic structure of quark propagator has
 interesting information about confinement, we will go to this topic again in the third subsection  of section 4.

The Euclidean rain-bow SDE can be projected into two coupled integral
equations for $A(\bar{p}^2)$ and $B(\bar{p}^2)$, the explicit expressions for those equations
can be found in Ref.\cite{WangWan,Wan96}. For simplicity, we
ignore the bar on $p$ and $k$ in the following notations.

\section{Bethe-Salpeter equation}
The BSE is a conventional approach in dealing with the two body
relativistic bound state problems \cite{BS51}. The quark theory of the mesons indicate that the mesons are
quark and anti-quark bound states. The precise knowledge about the quark structures of the mesons
will result in better understanding of their properties. In the following, we write down the
BSE for the pseudoscalar mesons,
\begin{eqnarray}
S^{-1}_{+}(q+\xi P)\chi(q,P)S^{-1}_{-}(q-(1-\xi)P)=\frac{16 \pi i}{3} \int \frac{d^4 k}{(2\pi)^4}\Gamma_\mu \chi(k,P)
\Gamma_\nu G_{\mu \nu}(q-k),
\end{eqnarray}
where $S(q)$ is the quark propagator function, $G_{\mu \nu}(k)$ is the gluon propagator, $P_\mu$ is the four-momentum
of the center of mass of the pseudoscalar mesons, $q_\mu$ is the relative four-momentum between the quark and anti-quark
in the pseudoscalar mesons, $\Gamma_{\mu}$ is the full vertex of quark-gluon, $\xi$ is the center of mass parameter
which can be chosen to between $0$ and $1$, and
 $\chi(q,P)$ is the Bethe-Salpeter wavefunction (BSW) of the bound state. In the limit $\Gamma_\mu=\gamma_\mu$,
we obtain the ladder BSE.

After we perform the Wick rotation analytically and continue  $q$ and $k$ into the Euclidean region,
the Euclidean pseudoscalar BSW $\chi(q,P)$ can be expanded in lorentz-invariant functions:
\begin{eqnarray}
\chi(q,P)=\gamma_5 \left[ iF_1(q,q \cdot P)+\gamma \cdot P F_2(q,q\cdot P)
+\gamma \cdot q F_3(q,q\cdot P)+i[\gamma \cdot q,\gamma \cdot P  ] F_4(q,q\cdot P) \right].
\end{eqnarray}
The BSW $F_{i}$ can be expressed in terms of the $SO(4)$ eigenfunctions, the Tchebychev polynomials
$T^{\frac{1}{2}}_{n}(\cos \theta)$,
\begin{eqnarray}
F_{i}(q,q\cdot P)=\sum_0^{\infty}F_{i}^{n}(q,P) q^n P^n T^{\frac{1}{2}}_{n}(\cos \theta),
\end{eqnarray}
where n $=$ even if i $=$ 1,2,4; n $=$ odd if n $=$ 3, $T^{\frac{1}{2}}_{n}(\cos \theta)=\cos(n \cos \theta)$ and
$\theta$ is the included angle between q and P.
In solving the coupled BSEs for $F_{i}^n$, it is impossible to solve an infinite series of coupled equations,
 we have to make truncations in one or the other ways in practical manipulations,
numerical calculations indicate that taking only $n=0,1$ terms can give satisfactory results,
\begin{eqnarray}
\chi(q,P)=\gamma_5 \left[ iF_1^{0}(q,P)+\gamma \cdot P F_2^{0}(q,P)
+\gamma \cdot q q\cdot P F_3^{1}(q,P)+i[\gamma \cdot q,\gamma \cdot P  ] F_4^{0}(q,P) \right].
\end{eqnarray}
For a thorough investigation of the solutions of the above BSWs, we must take full quark propagator and full quark-gluon
vertex,
again we are led to solve a divergent series of coupled SDEs and BSEs, truncations in one or the other ways for
the quark propagator and  quark-gluon vertex  are also  necessary. In this article, we take the bare vertex for both the SDE and BSE.

In solving the BSEs, it is important to translate the wavefunctions $F_{i}^{n}$ into the same dimension,
\begin{eqnarray}
F_{1}^{n}\rightarrow \Lambda^{2n}F_{1}^{n}, \, F_{2}^{n}\rightarrow \Lambda^{2n+1}F_{2}^{n}, \, F_{3}^{n}\rightarrow \Lambda^{2n+1}F_{3}^{n},
\,F_{4}^{n}\rightarrow \Lambda^{2n+2}F_{4}^{n}, \, q\rightarrow q/\Lambda, \, P\rightarrow P/\Lambda \, ,
\end{eqnarray}
where $\Lambda$ is some quantity of the dimension of mass.

Here we take a short digression to discussing the spectrum of the BSEs. In ideal conditions, a
precise solution of the BSE for the bound states of definite quantum numbers will reproduce the full
spectrum with the fundamental parameters of QCD, such as $SU(3)$ gauge invariance, quark masses, etc.
For example, the solutions of the BSE for  $0^{-+}$ mesons will result in a full pseudoscalar spectrum
for both the fundamental states and excited states such as $\pi^0,\pi(1300) \cdots$. However, the present conditions
are  far from the case, the truncated BSEs always result in a spectrum with more bound states (artifact) \cite{Metsch}. Moreover,
the spectrum is not the major subject which the present article concern. So in the article,
we take the masses of the pseudoscalar mesons as input parameters and make an investigation of the $\pi$ and B mesons BSWs for
both ladder approximation and bare quark propagator approximation.

The ladder BSE can be projected into four coupled integral equations in the following form,
\begin{eqnarray}
H(1,1)F_1^0(q,P)+H(1,2)F_2^0(q,P)+H(1,3)F_3^1(q,P)+H(1,4)F_4^0(q,P)&=&\int_0^{\infty}k^3dk \int_0^{\pi}\sin^2 \theta K(1,1), \nonumber \\
H(2,1)F_1^0(q,P)+H(2,2)F_2^0(q,P)+H(2,3)F_3^1(q,P)+H(2,4)F_4^0(q,P)&=&\int_0^{\infty}k^3dk \int_0^{\pi}\sin^2 \theta (K(2,2)+K(2,3)), \nonumber \\
H(3,1)F_1^0(q,P)+H(3,2)F_2^0(q,P)+H(3,3)F_3^1(q,P)+H(3,4)F_4^0(q,P)&=&\int_0^{\infty}k^3dk \int_0^{\pi}\sin^2 \theta (K(3,2)+K(3,3)), \nonumber \\
H(4,1)F_1^0(q,P)+H(4,2)F_2^0(q,P)+H(4,3)F_3^1(q,P)+H(4,4)F_4^0(q,P)&=&\int_0^{\infty}k^3dk \int_0^{\pi}\sin^2 \theta K(4,4),
\end{eqnarray}
the expressions of the $H(i,j)$ and $K(i,j)$ are cumbersome and neglected here, the interested readers can get the
word-version from the author.

Here we give some explanations about the expressions of $H(i,j)$ . The $H(i,j)$'s are functions of the quark's
Schwinger-Dyson functions (SDF) $A(q^2+\xi^2 P^2+\xi q \cdot P)$, $B(q^2+\xi^2 P^2+\xi q \cdot P)$,
 $A(q^2+(1-\xi)^2 P^2-(1-\xi) q \cdot P)$ and $B(q^2+(1-\xi)^2 P^2-(1-\xi) q \cdot P)$. The relative
four-momentum $q$ is a quantity in Euclidean space-time  while the center of mass four-momentum $P$ is a quantity
 in Minkowski space-time.
The present theoretical techniques  can not solve the SDE in Minkowski space-time, we have to expand $A$ and $B$
in terms of Taylor series of  $q \cdot P$,
\begin{eqnarray}
A(q^2+\xi^2 P^2+\xi q \cdot P)&=&A(q^2+\xi^2 P^2)+A(q^2+\xi^2 P^2)'\xi q \cdot P+\cdots, \nonumber \\
&\cdots&\nonumber \\
B(q^2+\xi^2 P^2+\xi q \cdot P)&=&B(q^2+\xi^2 P^2)+B(q^2+\xi^2 P^2)'\xi q \cdot P+\cdots.
\end{eqnarray}
The other problem is that we can not solve the SDE in the time-like region as the two
point gluon Green's function can not be exactly inferred from the $SU(3)$ color gauge theory
even in the low energy space-like region. In practical manipulations, we can extract the values of $A$ and $B$
from the space-like region smoothly to the time-like region with the polynomial functions.
To avoid possible violation with
confinement in sense of the appearance of pole masses $q^2=-m(q^2)$, we must be care in
the choice of polynomial functions \cite{Munczek91}.

Finally, we write down the normalization condition for the BSW,
\begin{eqnarray}
\int \frac{d^4q}{(2\pi)^4} \{ \bar{\chi} \frac{\partial S^{-1}(q+\xi P)} {\partial P_{\mu}}\chi(q,P) S^{-1}(q-(1-\xi)P)
+\bar{\chi} S^{-1}(q+\xi P) \chi(q,P) \frac{\partial S^{-1}(q-(1-\xi) P)} {\partial P_{\mu}} \}=2 P_{\mu},
\end{eqnarray}
where $\bar{\chi}=\gamma_4 \chi^+ \gamma_4$. We can substitute the expressions of the BSWs and SDFs into the above
equation and obtain the precise result, however, the expressions are cumbersome and  neglected here.
\subsection{Decay constants of pseudoscalar mesons}
The decay constants of the pseudoscalar mesons are defined by the following current-meson duality,
\begin{eqnarray}
i f_{\pi} P_\mu &=& \langle0|\bar{q}\gamma_\mu \gamma_5 q |\pi(P)\rangle, \nonumber \\
&=& \sqrt{N_c}\int Tr \left[\gamma_\mu \gamma_5\chi(k,P) \frac{d^4 k}{(2\pi)^4} \right],
\end{eqnarray}
here we use $\pi$ to represent the pseudoscalar mesons.

\subsection{Bethe-Salpeter equation with bare quark-gluon vertex and bare quark propagator}
In this subsection, we investigate the BSWs of the $\pi$ and B mesons with the quark-gluon vertex and
quark propagator are both taken to be bare,
\begin{eqnarray}
\Gamma_\mu=\gamma_\mu, \ \ \ S^{-1}(p)=i\gamma \cdot p +M,
\end{eqnarray}
where the effective mass $M$ is taken to be the constituent quark mass for the $u$, $d$ and $b$ quarks.
In this article, we take the effective mass $M$ as an input parameter.
Such a two point quark Green's function
can not embody dynamical chiral symmetry breaking and has a mass pole in the time-like region. However,
as a first step, we can study the BSWs for those pseudoscalar mesons with the crude approximation.
The algebraic expressions for the BSWs can be obtained easily with a simple substitution of $A(p)\rightarrow 1$ and $B(p)\rightarrow M$
in Eq.(14).
After solving the BSEs numerically by iterations, we plot the BSWs $F^0_1,F^0_2,F^1_3,F^0_4$ as functions of the relative four-momentum
$q$ for the $\pi$ meson and B meson, respectively. In this article, we take the $\pi$ meson BSWs explicitly shown in Fig.1 as
an example and neglect others for simplicity.
As the  values of the wavefunctions  $ F^1_3,F^0_4$ are tiny, we plot them perspicuously in another figure .
From those figures, we can  see that the first
two wavefunctions are dominating, $F^0_1,F^0_2 \gg F^1_3,F^0_4$, and all of the four wavefunctions are gaussian-type
and center around very small momentum i.e. near zero momentum which indicates  that the bound states must exist at the small momentum region or
in  other
words confinement occurs at the infrared region.
Based on the numerical values of the BSWs of the $\pi$ and B mesons, we can obtain the corresponding decay constants.
\begin{eqnarray}
f_\pi=134 MeV; & \ \ \  \ \ & f_B=164 MeV,
\end{eqnarray}
which are compatible with the experimental, lattice and QCD sum rule results, $f_{\pi}=130MeV (Exp)$ and $f_B\approx 150-210MeV (Latt,sumrule)$ \cite{Dominguez,Khodjamirian,Lattice}.
In calculation, the input parameters are $N=1.0 \Lambda $, $V(0)=-10.0 \Lambda$,
 $\rho=5.0\Lambda$, $M_{u}=M_{d}=530 MeV$, $M_b=5200 MeV$, $\Lambda=200 MeV$, $\varpi=1.3 GeV$ and  $\Delta=0.03 GeV^2$.
\subsection{Coupled rain-bow Schwinger-Dyson equation and ladder Bethe-Salpeter equation}
In this subsection, we explore the coupled equations of the rain-bow SDE and ladder BSE with the bare
 quark-gluon vertex for both the $\pi$ and B mesons. The algebraic expressions for those solutions are
 obtained already in section 3 and forepart of section 4, here we will not repeat the tedious routine.
In solving those equations numerically, the simultaneous iterations converge quickly to an unique value independent of
the choice of initial  wavefunctions. The final results for the SDFs and BSWs are plotted
as functions of the square momentum $q^2$.

 The quark-gluon vertex can be dressed through the solutions of the Ward-Takahashi identity or
 Salvnov-Taylor identity and taken to be the Ball-Chiu vertex and Curtis-Pennington vertex \cite{BallChiu,Curtis}. Although it is possible
to solve the SDE with the dressed vertex, our analytical results indicate that the expressions for the  BSEs with the dressed
vertex are cumbersome and not suitable for numerical iterations
\footnote{This observation is based on the authors's work in USTC. }.

In order to demonstrate the confinement of quark, we have to
study the SDF of the quark and prove
 that there no poles on the real timelike  $p^2$ axial.
So it is necessary to perform an analytic continuation of
the dressed quark propagator from Euclidean space into Minkowski
space $p_{4} \rightarrow ip_{0}$.
However, we have no knowledge of the singularity structure of
quark propagator in the whole complex plane. One can take an
alternative safe procedure and stay completely in Euclidean
space avoiding analytic continuations of the dressed
propagators \cite{Burkardt}. It is sufficient to take the Fourier transform
 with respect to the Euclidean time T
 for the scalar part $S_{s}$,
 \begin{eqnarray}
 S^{*}_{s}(T) & & =\int_{-\infty}^{+ \infty} \frac{dq_{4}}{2 \pi}
 e^{iq_{4}T}S_{s}, \nonumber \\
 & & =  \int_{-\infty}^{+ \infty} \frac{dq_{4}}{2 \pi} e^{iq_{4}T}
 \frac{B(q^2)}{q^2A^2(q^2)+B^{2}(p^2)}.
 \end{eqnarray}
 If S(p) had a pole at $p^2=-m^2$, the Fourier transformed
  $S^{*}_{s}(T)$ would fall off as $e^{-mT}$ for large T or
  $\log{S^{*}_{s}}=-mT$.

In our calculation, for large $T$, the values of $S^{*}_{s}$ is negative,
except occasionally a very small fraction positive values. We can express
$S^{*}_{s}$ as $|S^{*}_{s}|e^{i n\pi}$, $n$ is an odd integer.
$\log{S^{*}_{s}}=\log|{S^{*}_{s}}|+in \pi$.
If we neglect the imaginary part, we find that when the Euclidean time T is
large, there indeed exists a crudely approximated (almost flat) linear function with about zero slope for all the u, d (The curve for the d quark
 has the same behavior as the u quark in the limit of Isospin symmetry is exact ) and b quarks with
respect to T, which is shown in Fig.2. Here the word 'crudely' should be
understand in the linearly fitted sense, to be exact, there is no
linear function. However, such  fitted linear functions are hard
to acquire  physical explanation and the negative values for $S^{*}_{s}$ indicate  an explicit
violation of the axiom of reflection positivity \cite{Jaffee},
 in other words, the quarks are not physical observable i.e. confinement.

From Fig.3, we can see that for the $u$ and
$d$ quarks, the SDFs are greatly renormalized at small momentum region and the curves are steep at about $q^2=1 GeV^2$ which
indicates  an explicit dynamical chiral symmetry breaking, while at large $q^{2}$,
they take asymptotic behavior.
As for the $b$ quark, shown in Fig.4, the current mass is very large, the renormalization is more tender,
however, mass pole in the time-like region is also absent, which can be seen from Fig.2. At zero momentum, $m_u(0)=688 MeV $, $m_d(0)= 688 MeV $ and
$m_b(0)=4960 MeV $, which are compatible with the constituent quark masses. In fact,
 the connection of $m(p)$ to constituent masses is somewhat less direct
for the light quarks and is precise only for the heavy quarks. For heavy quarks,
$ m_{constituent}(p)=m(p=2m_{constituent}(p))$ , for light quarks ,
it only makes a crude estimation \cite{Politzer}. From the plotted BSWs (neglected here for simplicity), we can see that the BSWs for both
the $\pi$ and B mesons have the same type momentum dependence as the corresponding wavefunctions with the bare quark
propagators, however, the quantitative values are changed. The gaussian type BS wavefunctions which
center around small momentum indicate that the bound states exist only in the infrared region, in other words
confinement. Finally we obtain the values
for the decay constants of  the $\pi$ and B mesons,
\begin{eqnarray}
f_\pi=127 MeV; & \ \ \ \ \ \ \ & f_B=192 MeV,
\end{eqnarray}
which are compatible with the experimental, lattice and QCD sum rule results, $f_{\pi}=130MeV (Exp)$ and
$f_B\approx 150-210MeV (Latt, sumrule)$ \cite{Dominguez,Khodjamirian,Lattice}.
In calculation, the input parameters are $N=1.0 \Lambda $, $V(0)=-11.0 \Lambda$,
 $\rho=5.0\Lambda$, $m_{u}=m_{d}=6 MeV$, $m_b=4700 MeV$, $\Lambda=200 MeV$, $\varpi=1.6 GeV$ and  $\Delta=0.04 GeV^2$.

From the variations of the values for the decay constants of both the $\pi$ and B mesons, we can estimate  that
the full vertex approximation will not change those values greatly.

\section{conclusion and discussion}
In this article, we investigate the under-structures of the $\pi$ and B mesons in the framework of
the Bethe-Salpeter equation with the confining effective potential (infrared modified flat bottom potential).
In bare quark-gluon vertex approximation, we obtain the
algebraic expressions for the solutions of the coupled rain-bow SDE and ladder
BSE for those mesons. At the first step, we neglect the rain-bow SDE,
take the bare quark propagator and solve the BSE numerically alone. Although the bare quark propagator can not embody dynamical chiral symmetry breaking and
has a mass pole in the time-like region, it can give reasonable results for the values of the
decay constants $f_\pi$, $f_B$ compared with the values of the experimental data and other theoretical calculations, such as
lattice simulations  and QCD sum rules. In calculation, we obtain the BSWs for the $\pi$ and B mesons, which
center in the small momentum region, are compatible with confinement.
Secondly, we explore the coupled equations of the rain-bow SDE and ladder BSE with the bare
quark-gluon vertex for those mesons.
The quark-gluon vertex can be dressed through the solutions of the Ward-Takahashi identity or
 Salvnov-Taylor identity and taken to be the Ball-Chiu vertex and Curtis-Pennington vertex, however,
a consistently numerical manipulation is unpractical. After we solve the coupled rain-bow SDE and ladder
BSE numerically, we obtain  both the SDFs and BSWs for both the $\pi$ and B mesons. The  SDFs for the $u$ and
$d$ quarks are greatly renormalized  at small momentum region and the curves are steep at about $q^2=1 GeV^2$ which
indicates an explicitly dynamical chiral symmetry breaking. After we take Euclidean time fourier transformation about
the quark propagator, we can find that there is no mass pole in the time-like region and obtain satisfactory
result about confinement. As for the $b$ quark, the current mass is very large, the renormalization is more tender,
however,  mass pole in the time-like region is also absent. The BSWs for both
the $\pi$ and B mesons have the same type momentum dependence as the corresponding wavefunctions with the
bare quark propagators,
 however, the quantitative values are changed and the corresponding values for the decay constants
$f_\pi$ and $f_B$ are changed, but not greatly.
 We can estimate  that the full vertex approximation will not change those values greatly.
Once the SDFs and BSWs for both the $\pi$ and B mesons are known, we can use them to investigate a lot of important
quantities in the B meson decays, such as $B-\pi$, $B-K$, $B-\rho$ former factors, Isugar-Wise functions, strong coupling constants,
 etc.

\section*{Acknowledgment}

The author (Z.G.Wang) would like to thank National
Postdoctoral Foundation for financial support. The author will also thanks Dr.Gogohia for helpful discussion
about the solutions of SDE.

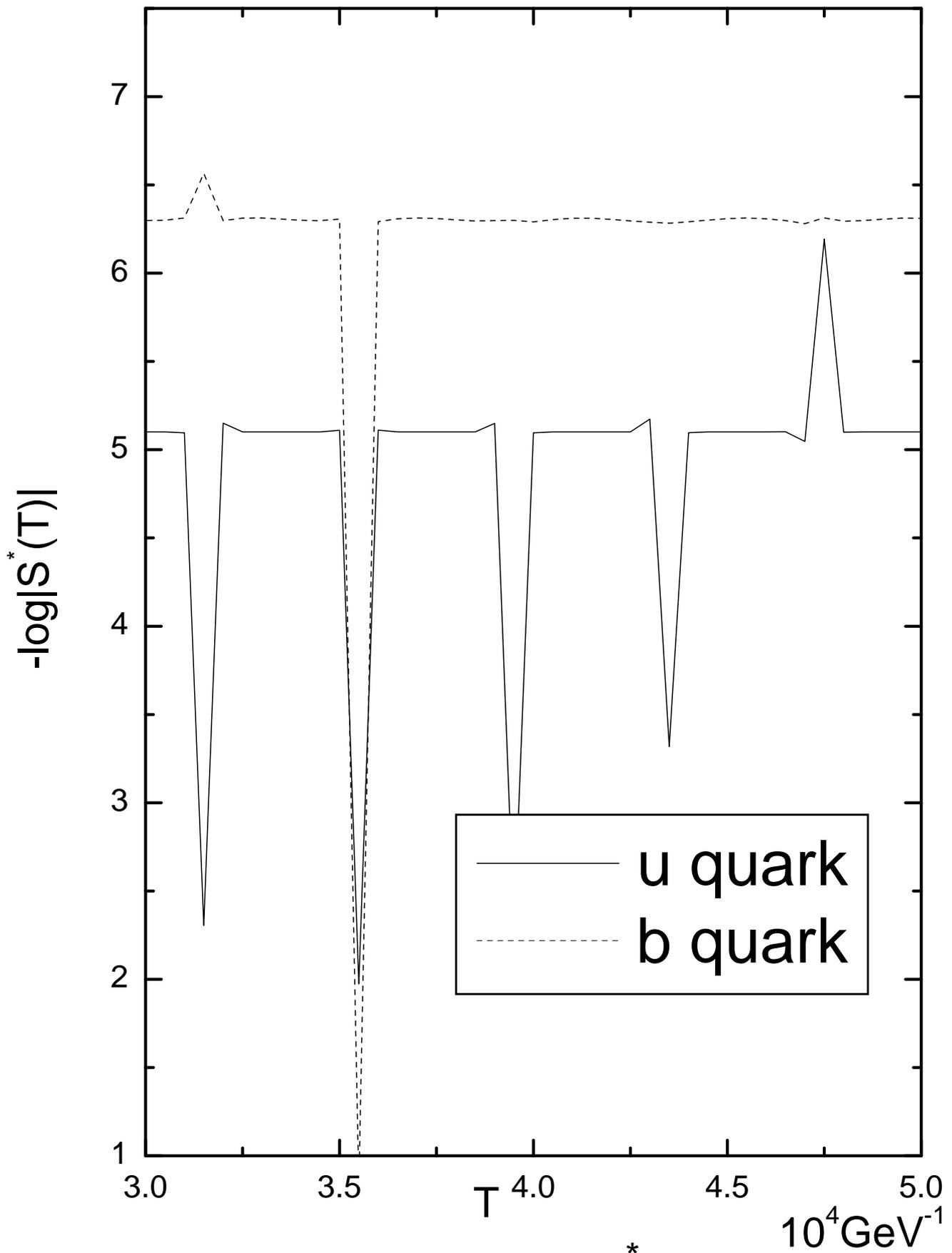
\end{document}